\documentclass[lettersize,journal]{IEEEtran}
\usepackage{amsmath,amsfonts}
\usepackage{algorithmic}
\usepackage{algorithm}
\usepackage{array}
\usepackage[caption=false,font=normalsize,labelfont=sf,textfont=sf]{subfig}
\usepackage{textcomp}
\usepackage{stfloats}
\usepackage{url}
\usepackage{verbatim}
\usepackage{graphicx}
\usepackage{cite}
\usepackage{amssymb}
\usepackage{cuted}
\usepackage{epsfig}
\usepackage{flushend}
\usepackage{color}
\usepackage{soul}
\begin{document}
	\title{ RIS-Aided Dual-Polarized MIMO: 
		How Large a Surface is Needed to Beat Single Polarization?
	}

	\author{Zizhou Zheng, Huan Huang, Hongliang Zhang, and A. Lee Swindlehurst
		\thanks{
			Z. Zheng is with the School of Electronics, Peking University, Beijing 100871, China, and also with School of Information Science and Engineering, Southeast University, Nanjing, China (e-mail: 213211229@seu.edu.cn). 
			
			H. Huang is with the School of Electronic and Information Engineering, Soochow University, Suzhou, Jiangsu 215006, China (hhuang1799@gmail.com). 
			
			H. Zhang is with the School of Electronics, Peking University, Beijing 100871, China (e-mail: hongliang.zhang92@gmail.com). 
			
			A. Lee Swindlehurst is with the Center for Pervasive Communications and Computing, University of California, Irvine, CA 92697 USA (e-mail: swindle@uci.edu).
			
			\emph{Corresponding authors: Hongliang Zhang}
		}	
	}

	\markboth{}%
	{Shell \MakeLowercase{\textit{et al.}}: A Sample Article Using IEEEtran.cls for IEEE Journals}
	\maketitle
	
	\begin{abstract}
		Dual-polarized (DP) multiple-input-multiple-output (MIMO) systems have been widely adopted in commercial mobile wireless communications. Such systems achieve multiplexing and diversity gain by exploiting the polarization dimension. However, existing studies have shown that the capacity of DP MIMO may not surpass that of single-polarized (SP) MIMO systems due to the cross-polarization coupling induced by the propagation environment.
		In this letter, we employ reconfigurable intelligent surfaces (RISs) to address this issue and investigate how large the surface should be to ensure a better performance for DP MIMO. Specifically, we first derive the capacities of DP and SP MIMO systems with an RIS, and then study the influence of the RIS size on the system capacity. Our analyses reveal how to deploy the RIS in a DP MIMO scenario.
	\end{abstract}
	
	\section{Introduction}	
	\IEEEPARstart{P}{olarization} refers to the spatial variation of the orientation of the wave's electric field vector. Measurements have shown that vertically and horizontally polarized electromagnetic (EM) waves exhibit nearly independent propagation characteristics \cite{4349833}. Consequently, a compact antenna array that integrates orthogonal polarized antennas can be realized to provide multiplexing and diversity gain in the polarization domain. Benefiting from the performance gains in the polarization domain, dual-polarized (DP) multiple-input-multiple-output (MIMO) systems have been widely adopted in commercial mobile communication systems \cite{6163590,6476877}.
	
	However, DP MIMO may not outperform single-polarized (SP) systems under certain conditions. In practice, some scatterers in the propagation environment can cause a cross-polarization effect that alters the polarization state of the EM waves and produces cross-polarization interference. In such cases, DP MIMO performance may not exceed that of SP systems. This issue is further worsened with low transmission power, as the degrees of freedom in the polarization domain are not effectively utilized \cite{4525659}. This implies that the benefits of DP MIMO systems rely on a favorable propagation environment.
	
	Recently, reconfigurable intelligent surfaces (RISs) have been introduced, and polarized RISs have shown potential in addressing the above issues \cite{10555049,9110889}. A DP RIS can adjust the phase shifts of impinging signals across both polarizations, thereby shaping the polarization-dependent propagation environment and enhancing the link quality.
	\textcolor{black}{Some studies on the DP RIS have been reported \cite {9497725,10256051,9440813, 9900387}. In \cite {9497725}, the authors conducted a performance analysis of a DP RIS-assisted communication system with imperfect polarization isolation. The work of \cite {10256051} proposed a broad-beam generation method for polarized RISs by adjusting its phase shifts in orthogonal polarization dimensions. A multi-user  MIMO non-orthogonal multiple access network aided by a DP RIS was reported in \cite{9440813}. In \cite{9900387}, the authors proposed a dual-polarized cooperative RIS-assisted MIMO device-to-device communication scheme.}
	
	Existing studies on DP RIS-aided MIMO systems focus on the phase shift design for RIS with fixed dimensions. However, for certain RIS sizes, superior DP capacity may not be possible. Therefore, it is of interest to study the DP RIS size required to outperform SP MIMO. In this letter, we aim to answer this question by analyzing the impact of the RIS size on the system capacity. The main contributions are summarized as follows:
	\begin{itemize}
		\item We derive the capacities of RIS-aided DP and SP MIMO systems, accounting for polarization leakage, transmission power, and RIS size. 
		\item Based on this, we derive the RIS size required to outperform the SP MIMO capacity.
		\item We further analyze the impact of the transmission power and polarization leakage on the required RIS size. 
	\end{itemize}
	
	The rest of this paper is organized as follows. In Section II, we introduce the DP and SP system models for RIS-assisted communications. In Section III, we derive the capacities of RIS-assisted DP and SP MIMO systems and find the RIS size required for DP systems to have a capacity superior to the SP case.  Numerical results in Section IV validate our analysis. Finally, conclusions are drawn in Section V.
	
	\section{System Model}
	\subsection {Scenario Description}
	\begin{figure}[t]
		\centering
		\includegraphics[width=3in]{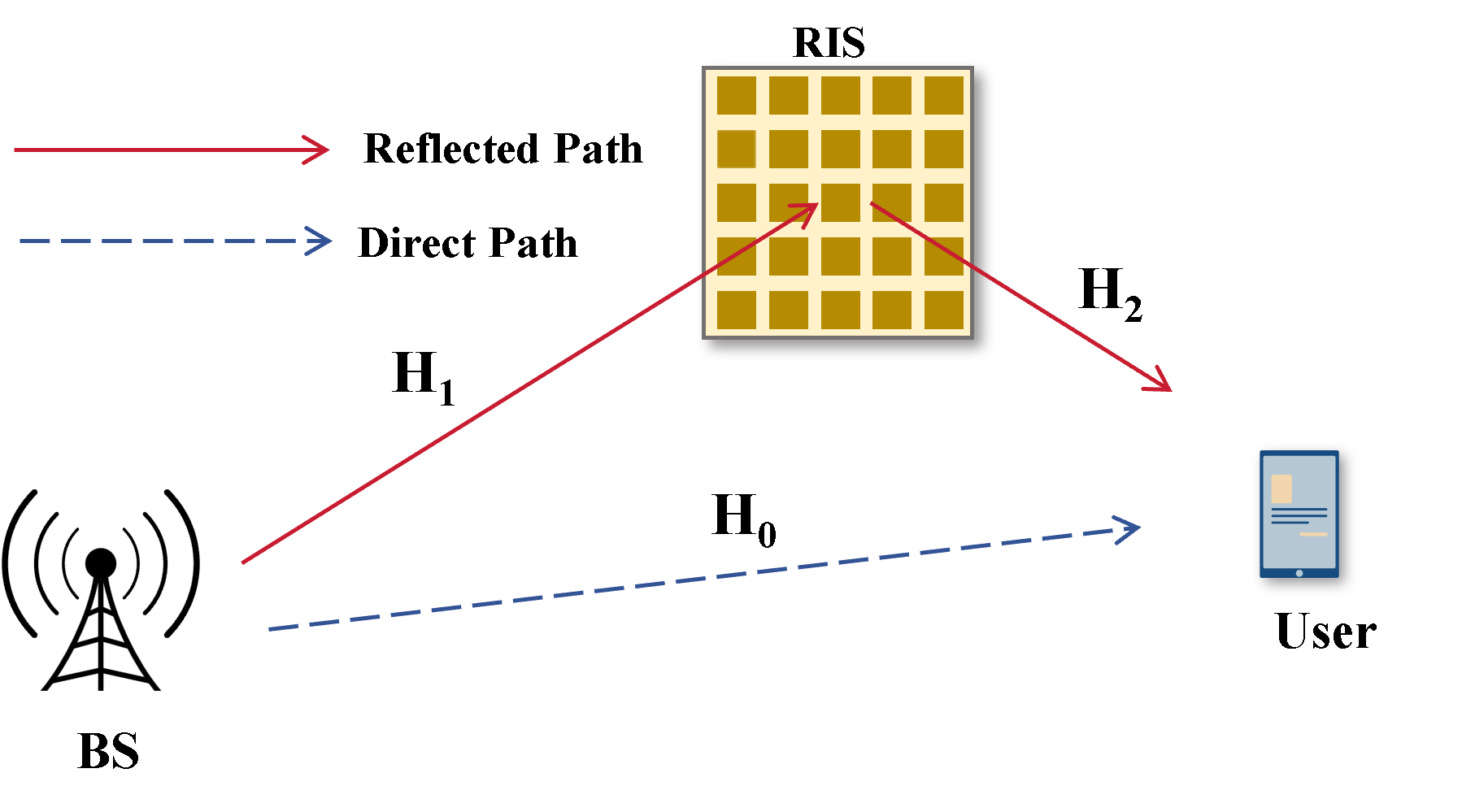}
		\caption{\textcolor{black}{System model for an RIS-aided MIMO downlink system.}}
		\label{fig_1}
	\end{figure}
	We consider RIS-aided DP and SP MIMO systems, as shown in Fig. {\ref {fig_1}}.  To conduct a fair comparison of capacity, the base station (BS) is equipped with either 2$N_t$ SP antennas or $N_t$ DP antennas, where each DP antenna can be regarded as a pair of co-located vertically and horizontally polarized antennas. Similarly, the user and the RIS are equipped with two SP antennas and 2$L$ SP elements in the SP system, or one DP antenna and $L$ DP elements in the DP system, respectively. From a modeling viewpoint,  each DP element can be considered as two co-located SP elements that separately interact with the orthogonally polarized components of the signal. \textcolor{black}{Consequently, we define the phase shift matrix of the DP RIS as
	\begin{equation}\label{Phi}
		\mathbf{\Phi}=\left[\begin{array}{cc}
			\mathbf{\Phi}_{0} & \bf 0 \\
			\bf 0 & \boldsymbol{\Phi}_{1}
		\end{array}\right]^{2 L \times 2 L} ,
	\end{equation}}
	\textcolor{black}{where ${\bf \Phi}_p=\operatorname{diag}\left\{e^{j \phi_{p,1}}, \ldots, e^{j \phi_{p, L}}\right\}$, $p\in \{0,1\}$ denotes the phase shift in polarization $p$,} and the phase shift matrix of the SP RIS is denoted by ${\bf \Phi}=\operatorname{diag}\left\{e^{j \phi_{1}}, \ldots, e^{j \phi_{2L}}\right\}$. 
	\subsection{Channel Model}
	The initial polarization state of the EM wave at transmission changes during propagation. For an SP channel, this change simply results in a lower received signal-to-noise ratio (SNR). However, for a DP channel, it leads to cross-polarization interference. To describe the ability of a channel to discriminate between two polarizations, we introduce the channel cross-polarization discrimination (XPD), defined as
	\begin{equation}
		{\rm XPD}=\frac{\mathbb E{\{||{{\mathbf{h}}}_{00}||^2\}}}{\mathbb E{\{||{ {\mathbf{h}}}_{10}||^2\}}}=\frac{\mathbb E{\{||{{\mathbf{h}}}_{11}||^2\}}}{\mathbb E{\{||{{\mathbf{h}}}_{01}||^2\}}},
	\end{equation}
	where $\mathbf{h}_{pq} \in {\mathbb C}^{1 \times N_t}$ represents the channel from polarization $q$ to polarization $p$, and $p,q \in \{0, 1\}$ denote the vertical and horizontal polarization, respectively.
	
	The channels from the BS to the user, from the BS to the RIS, and from the RIS to the user are modeled by ${\bf H}_0 \in {\mathbb C}^{2 \times 2N_t} $, ${\bf H}_1\in {\mathbb C}^{2L \times 2N_t}$, and ${\bf H}_2 \in {\mathbb C}^{2 \times 2L}$, respectively.
	\textcolor{black}{We assume that the instantaneous channel state information
		(CSI) of the above channels is unknown at the BS, and their statistical CSI distributions are identical.
		Thus, the power should be uniformly allocated to all $2N_t$ antennas.}
	
	\textit{ 1) DP channels:} For the case of Rayleigh fading,  ${\bf H}_0$ is expressed as
	\begin{equation}
		{\bf H}_0=\sqrt{\beta_{0}}\left[\begin{array}{ll} 
			\tilde{\mathbf{h}}_{00} & \tilde{\mathbf{h}}_{01} \\
			\tilde{\mathbf{h}}_{10} & \tilde{\mathbf{h}}_{11}
		\end{array}\right],
	\end{equation}
   \textcolor{black}{where $\beta_{0}$ is the path loss and the elements of $\tilde{\mathbf{h}}_{pq} \in \mathbb C^{1 \times N_t}$ follow a complex Gaussian distribution with zero mean.} The variances of the co-polarized and cross-polarized components are respectively equal to $1-\alpha$ and $\alpha$. Thus, the XPD of ${\bf H}_0$ is denoted by
	\begin{equation}
		{\rm XPD}_{\rm N}=\frac{\mathbb E{\{||{\tilde {\mathbf{h}}}_{00}||^2\}}}{\mathbb E{\{||{\tilde {\mathbf{h}}}_{10}||^2\}}}=\frac{\mathbb E{\{||{\tilde {\mathbf{h}}}_{11}||^2\}}}{\mathbb E{\{||{\tilde {\mathbf{h}}}_{01}||^2\}}}=\frac{1-\alpha}{\alpha}, 
	\end{equation}
	where the variance $\alpha \in (0,1]$ corresponds to the power leaked into the cross-polarized components. Since the channel cannot introduce more power into the transmitted signal \cite {4525659},  the power of the co-polarized components should be $1-\alpha$, which ensures a fair comparison between the DP and SP MIMO systems.
	
	The ${2 L \times 2 N_{t}}$ channel ${\bf H}_{1}$ from the BS to the RIS is modeled as
	\begin{equation}
		\begin{aligned}\label {H1}
			&	\mathbf{H}_{1}=\sqrt{\beta_{1}}\left[\begin{array}{ll}
				\mathbf{H}_{1,00} & \mathbf{H}_{1,01} \\
				\mathbf{H}_{1,10} & \mathbf{H}_{1,11}
			\end{array}\right], \\
			&	=\sqrt{\beta_{1}}\left[\begin{array}{ll}
				\sqrt{1-\alpha_{f_{1}}} e^{j \psi_{1,00}}\widehat{\mathbf{H}}_{1} & \sqrt{\alpha_{f_{1}}} e^{j \psi_{1,01}}\widehat{\mathbf{H}}_{1} \\
				\sqrt{\alpha_{f_{1}}} e^{j \psi_{1,10}}\widehat{\mathbf{H}}_{1} & \sqrt{1-\alpha_{f_{1}}} e^{j \psi_{1,11}}\widehat{\mathbf{H}}_{1}
			\end{array}\right],
		\end{aligned}
	\end{equation}
	where $\beta_{1}$ presents the path loss, $\alpha_{f_1}$ denotes the XPD factor, and $\psi_{1,pq}$ are the phase shifts for the corresponding polarization pair. Assuming a uniform linear array (ULA) at both the BS and RIS, the shared channel can be written as $\widehat{\mathbf{H}}_{1}=\mathbf{a}_{L}\left(\theta_{\mathrm{AoA}, 1}\right) \mathbf{a}_{N_{t}}^{\mathrm{T}}\left(\theta_{\mathrm{AoD}, 1}\right)$, where
	\begin{equation}
		\mathbf{a}_{N}(\theta)=\left[1, e^{j 2 \pi \frac{d}{\lambda} \sin \theta}, \cdots, e^{j 2 \pi(N-1) \frac{d}{\lambda} \sin \theta}\right]^{\rm T}
	\end{equation}
	denotes the steering vector of an $N$-element ULA, $d$ is the antenna separation, $\lambda$ is the wavelength, $\theta_{\mathrm{AoD}, 1}$ is the angle of departure (AoD) at the BS, and $\theta_{\mathrm{AoA}, 1}$ is the angle of arrival (AoA) at the RIS.
	
	Similarly, the $2 \times 2L$ channel ${\bf H}_{2}$ from the RIS to the user is denoted as
	\begin{equation}\label{H2}
		\begin{aligned}
			&\mathbf{H}_{2}= \sqrt{\beta_{2}}\left[\begin{array}{ll}
				\sqrt{1-\alpha_{f_{2}}} e^{j \psi_{2,00}}\widehat{\mathbf{H}}_{2} & \sqrt{\alpha_{f_{2}}} e^{j \psi_{2,01}}\widehat{\mathbf{H}}_{2} \\
				\sqrt{\alpha_{f_{2}}} e^{j \psi_{2,10}}\widehat{\mathbf{H}}_{2} & \sqrt{1-\alpha_{f_{2}}} e^{j \psi_{2,11}}\widehat{\mathbf{H}}_{2}
			\end{array}\right],
		\end{aligned}
	\end{equation}
	where $\beta_{2}$ is the path loss, $\alpha_{f_2}$ denotes the XPD factor, and $\psi_{2,pq}$ are the phase shifts for the corresponding polarization pair. Note that the two differently polarized antennas at the user are co-located. Defining the AoD at the RIS as $\theta_{\mathrm{AoD}, 2}$, the shared component is denoted by $\widehat{\mathbf{H}}_{2}=\mathbf{a}_{L}^{\mathrm{T}}\left(\theta_{\mathrm{AoD}, 2}\right)$.

	\textit {2) SP channels:} The impact of polarization leakage in the SP MIMO system differs from the DP case when we assume all the antennas are identically polarized. Consequently, the elements of ${\bf H}_0$ follow $\mathcal{C N}(0$, $1-\alpha)$, and ${\bf{H}}_{1}$, ${\bf{H}}_{2}$ then reduce to
	\begin{equation}
		\begin{aligned}
			&\mathbf{H}_{1}=\sqrt{\beta_{1}} \sqrt{1-\alpha_{f_{1}}} e^{j \psi_{1}}\mathbf{a}_{2 L}\left(\theta_{\mathrm{AoA}, 1}\right) \mathbf{a}_{2 N_{t}}^{\mathrm{T}}\left(\theta_{\mathrm{AoD}, 1}\right),\\
			&\mathbf{H}_{2}=\sqrt{\beta_{2}} \sqrt{1-\alpha_{f_{2}}} e^{j \psi_{2}} \mathbf{a}_{2}\left(\theta_{\mathrm{AoA}, 2}\right) \mathbf{a}_{2 L}^{\mathrm{T}}\left(\theta_{\mathrm{AoD}, 2}\right),
		\end{aligned}
	\end{equation} 
	where $\psi_{1}, \psi_{2}$ are the corresponding polarization phase shifts.
	\section{Channel Capacity Analysis}
	\textcolor{black}{Since the transmission power is uniformly allocated to all antennas, the signal received  by the user is expressed as}
	\begin{equation}
		{\bf y}=\sqrt{\frac{P}{2N_t}}\left({\bf H}_2{\bf \Phi}{\bf H}_1 +{\bf H}_0\right){\bf s}+{\bf n},
	\end{equation}
	where $P$ is the transmission power, ${\bf s} \in {\mathbb C}^{2N_t \times 1} \sim ~{\cal C}{\cal N}~( 0,~{\bf I}_{2N_t})$ is the transmit signal, and ${\bf n}\sim ~{\cal C}{\cal N}~( 0,~{\bf I}_{2})$ denotes additive white Gaussian noise (AWGN).
	
	\textcolor{black}{We define ${\bf H}={\bf H}_2{\bf \Phi}{\bf H}_1 +{\bf H}_0$.} Then, the ergodic capacity of the $2 \times 2N_t$ MIMO channel is calculated by
	\begin{equation}
		\begin{aligned}
			{\rm C}&=\mathbb E\left\{\log _{2} \operatorname{det}\left(\mathbf{I}_{2}+\frac{ P}{2 N_{t}} \mathbf{H H}^{\rm H}\right)\right\} \\
			& =\mathbb E\left\{\log _{2} \prod_{i=1}\left(1+\frac{ P}{2 N_{t}} \lambda_{i}\right)\right\}\le {\rm log_2}W,
		\end{aligned}
	\end{equation}
	where 
	\begin{equation}
		\begin{aligned}
			{ W}&=\mathbb E\left\{\prod_{i=1}\left(1+\frac{ P}{2 N_{t}} \lambda_{i}\right)\right\} \\
			&=1+\frac{ P}{2 N_{t}} \mathbb{E}\left\{\sum_{i=1}^2 \lambda_{i}\right\}+\left(\frac{P}{2 N_{t}}\right)^{2} \mathbb{E}\left\{\prod_{i=1}^2 \lambda_{i}\right\},\label{W}
		\end{aligned}
	\end{equation}
	and $\lambda_1, \lambda_2$ are the eigenvalues of $\bf H H^{\rm H}$. 
	
	Considering the case without an RIS, i.e., there is no reflected link, the co-polarization and cross-polarization components of the DP channel have gains equal to $1-\alpha$ and $\alpha$, respectively. For the SP channel, only co-polarization components exist with the gain $1-\alpha$. When $\alpha = 0$, the DP channel is full-rank while the SP channel is rank-deficient. Consequently, when $\alpha \in (0,1]$, the SP channel tends to have a much larger $\lambda_{1}$ and a $\lambda_{2}$ near zero, leading to a bigger $\sum_{i=1}^2 \lambda_{i}$ but smaller $\prod_{i=1}^2 \lambda_{i}$. Therefore, when $P$ is small, the DP capacity is lower than the SP capacity. 
	
	\textcolor{black}{An upper bound for the RIS-aided DP and SP capacities is obtained by calculating $\log_2W$ with $W$ defined in (\ref {DP_op}), (\ref {SP_op}), and the variables $a_{pq}, b_{pq}, c_{pq}, r(P)$ do not depend on the RIS. Definitions and derivations are given in Appendix. \ref {Apb}.}
	\begin{figure}[t]
		\centering	
		\includegraphics[width=3in]{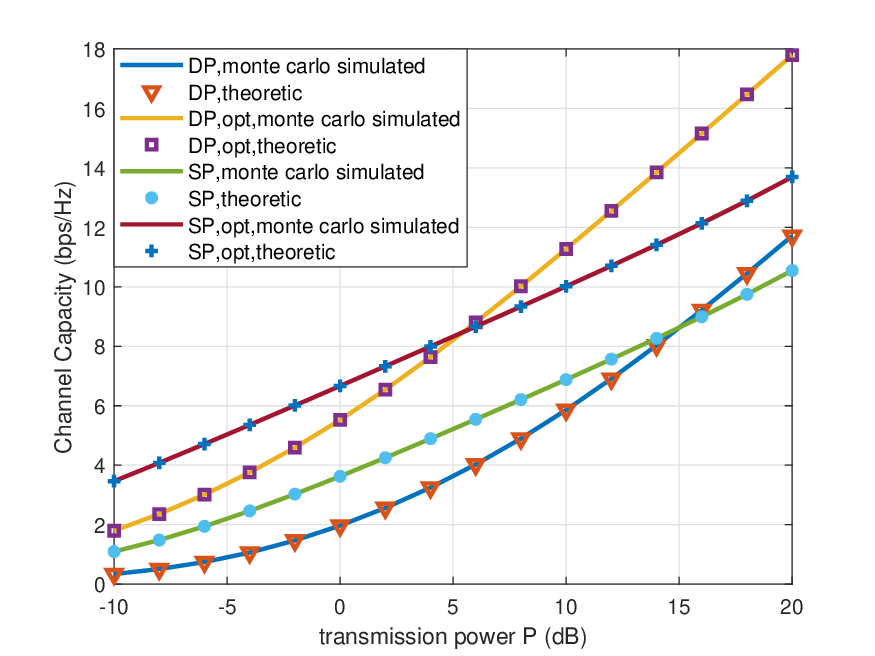}
		\caption{Capacity vs. transmission power $P$, $\beta_0=\beta_1=\beta_2=0.2$.}
		\label{fig2}
	\end{figure}

	\begin{figure*}[hb] 
		\centering 
		\hrulefill 
		\vspace*{1pt} 
		\textcolor{black}{	\begin{flalign}\label{DP_op}
		&	\begin{array}{l}
				{\rm W}^{\rm max}_{\rm DP}=1+P\left(\beta_{1} \beta_{2} L^{2}+\beta_{0}\right)+|r(P)| L^{2} \\
				+\frac{P^2}{4 N_{t}}\left(\begin{array}{l}
					\left|a_{00} b_{11}+b_{00} a_{11}-a_{01} b_{10}-a_{10} b_{01}\right|^{2}N_{t} L^{4}+ \beta_{0}^{2}\left(N_{t}-2 \alpha+2 \alpha^{2}\right)+2N_tL^2\beta_{0}\beta_{1}\beta_{2}\\
					- \beta_{0}\alpha L^{2}\left(\left|a_{00}\right|^{2}+\left|a_{11}\right|^{2}+\left|b_{00}\right|^{2}+\left|b_{11}\right|^{2}\right)-\beta_{0}(1-\alpha)L^{2}\left(\left|a_{10}\right|^{2}+\left|a_{01}\right|^{2}+\left|b_{10}\right|^{2}+\left|b_{01}\right|^{2}\right) \\
				\end{array}\right),
			\end{array}& 
		\end{flalign}} \textcolor{black}{
		\begin{flalign}\label{SP_op}
			&\begin{array}{l}
				{\rm W}^{\rm max}_{\rm SP}	=1+P\left(2 \beta_{0}(1-\alpha)+8 \beta_{1} \beta_{2}\left(1-\alpha_{f_{1}}\right)\left(1-\alpha_{f_{2}}\right) L^{2}\right) \\
				+\frac{P^2}{2 N_{t}}\left(8\left(2 N_{t}-1\right) \beta_{0} \beta_{1} \beta_{2}(1-\alpha)\left(1-\alpha_{f_{1}}\right)\left(1-\alpha_{f_{2}}\right) L^{2}+\left(2 N_{t}-1\right) (1-\alpha)^{2}\beta_{0}^{2}\right).
			\end{array}&
		\end{flalign}}
	\end{figure*}
	
	It can be observed that the DP capacity is proportional to $L^4$, while the SP capacity is proportional to $L^2$. When the RIS is sufficiently large, the DP capacity exceeds the SP capacity. 

	\textcolor{black}{Thus, there is a specific RIS size threshold beyond which dual polarization performs better than single polarization.}
		
	Next, we will discuss how large the RIS must be for the DP capacity to be larger than that for the SP case. We can obtain the threshold size  $L_{\rm req}$ by solving the following equation
	\begin{equation}\label{sul}
		{\rm W}^{\rm max}_{\rm DP}-{\rm W}^{\rm max}_{\rm SP}=d_{1} L_{\rm req}^{4}+d_{2} L_{\rm req}^{2}+d_{3}=0,
	\end{equation}
	where $d_1, d_2, d_3$ are given in (\ref {d}).

	\textcolor{black}{\textit{Proposition 1:} The ${\rm XPD}_{\rm N}$ factor $\alpha$ and $d_3$ satisfy:}
	\begin{equation}\label {161}
		d_3(2\alpha-1) \ge 0.
	\end{equation}
	
	\textit{Proof:} Details are given in Appendix. \ref{Apa}.
	
	In \cite {4657346}, measurements have been carried out demonstrating that ${\rm XPD}_{\rm N}=\frac{1-\alpha}{\alpha}$ is usually larger than 0 dB, which implies $\alpha < 0.5$. Thus, we consider the scenario where $d_3<0$. Since $d_1>0$, there is only one positive solution to (\ref {sul}), which then corresponds to the number of RIS elements required to ensure a better DP capacity:
	\begin{equation}\label{leq}
	\textcolor{black}{L_{\rm {req}}=\left \lceil \sqrt{\frac{-d_{2}+\sqrt{d_{2}^{2}-4 d_{1} d_{3}}}{2 d_{1}}} \right \rceil,}
   \end{equation}
 	\textcolor{black}{where $\lceil *\rceil$ denotes the ceiling function.}
	
	 \textcolor{black}{\textit{Proposition 2:} There is a lower limit on the required RIS size to ensure better DP performance, even if $P$ is infinite.}
	
	\textit{Proof:} When $P$ is large enough, $d_1, d_2$, and $d_3$ are mainly influenced by the terms involving $P^2$, so the highest order of $P$ in both the numerator and denominator of expression (\ref {leq}) is quadratic.
	Thus, (\ref {leq}) converges to a constant as $P \rightarrow\infty$, and we obtain  $\textit{Proposition 2}$.	
	
	\section{Simulation Results}
	In this section, we verify the derivation of the capacity with the optimal phase shift design, the RIS size required to outperform single polarization, and then investigate the impact of the transmission power and polarization leakage on the required size. The details of the simulation parameters are as follows: path losses $\beta_0 = 0.2$, $\beta_1 = 0.2$, $\beta_2 = 0.2$,  number of BS antennas $2N_t = 8$, number of RIS elements $L = 20$, XPD factors $\alpha = 0.14$, $\alpha_{f_1} = 0.1$, $\alpha_{f_2} = 0.13$, polarization phase shifts $\psi_{1,pq}=\psi_{2,pq}=0$, SP channel phase shifts $\psi_{1}=\psi_{2}=0$.
	
	\begin{figure}[t]
		\centering	
		\includegraphics[width=3in]{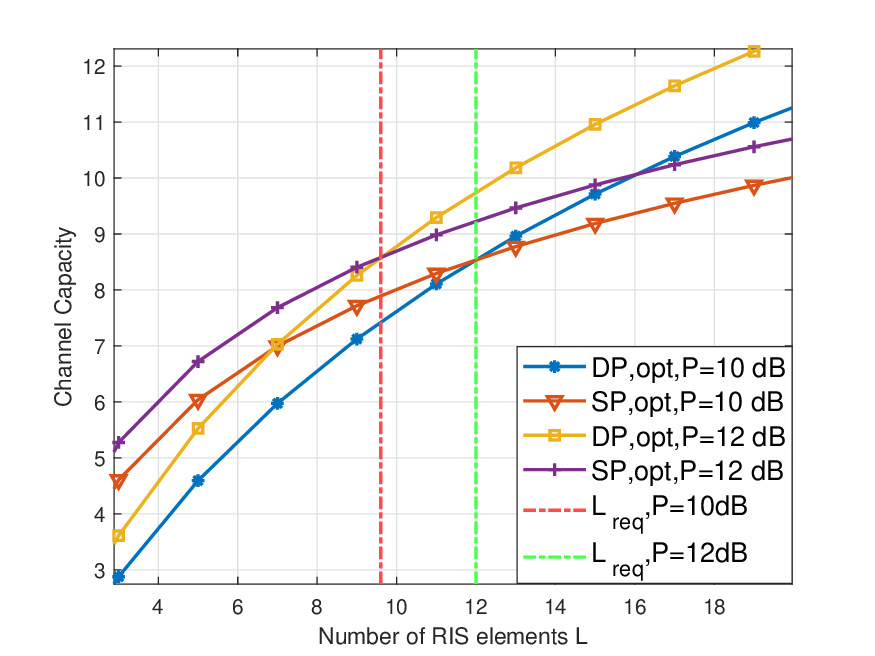}
		\caption{The verification of the required number of elements with $P= 10$, $12$ $\rm dB$.}
		\label{fig5}
		
		\centering	
		\includegraphics[width=3in]{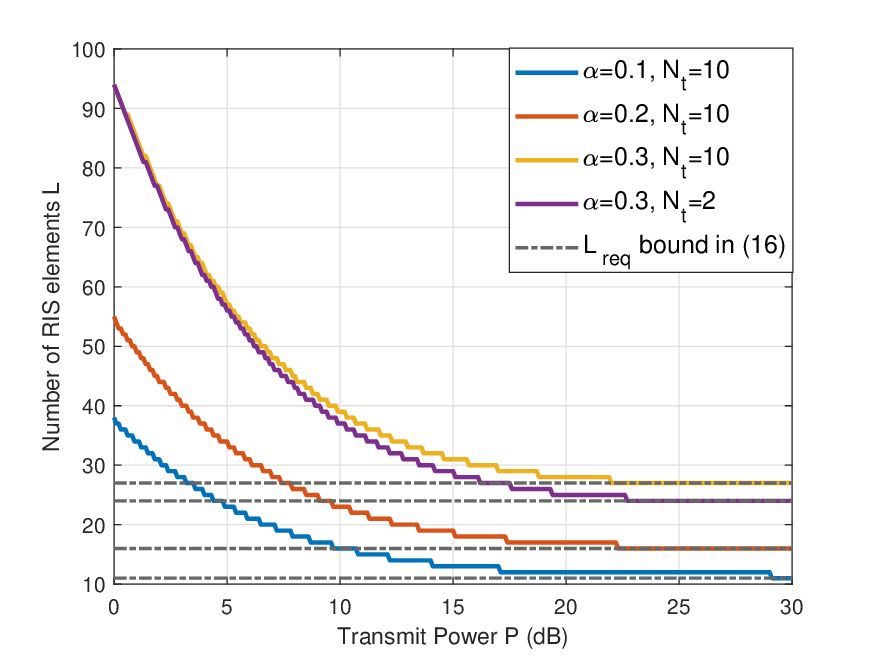}
		\caption{\textcolor{black}{Transmission power vs. RIS size, $\psi_{1,pq}=\psi_{2,pq}=0$, $\alpha=\alpha_{f_1}=\alpha_{f_2}$.}}
		\label{fig4}
	\end{figure}
	\textcolor{black}{In Fig. \ref{fig2}, we plot the capacity versus the transmission power $P$ for both the cases of random and optimal phase shifts. We observe that the theoretic results for $\log_2 (W^{\rm max})$ match the Monte Carlo simulations for $C$ in (10) very closely.}
	DP capacity increases faster with the transmission power since it has a larger $\prod_{i=1}^2 \lambda_{i}$, as discussed for (\ref {W}). In addition, we note that for this RIS size, when $P$ is lower than 5 $\rm dB$, the optimal phase shift design can not beat single polarization. This is consistent with our previous discussion. Therefore, it is crucial to investigate the impact of RIS size on the capacity.
	
	\begin{figure*}[hb] 
		\hrulefill
		\centering 
		\begin{subequations}\label{d}
			\begin{align}
				&d_{1}=\left(\frac{P}{2}\right)^{2}\left|a_{00} b_{11}+b_{00} a_{11}-a_{01} b_{10}-a_{10} b_{01}\right|^{2}, \\
				&d_{2}=|r(P)|+P \beta_{1} \beta_{2} -8 P \beta_{1} \beta_{2}\left(1-\alpha_{f_{1}}\right)\left(1-\alpha_{f_{2}}\right)-4\frac{P^2}{N_{t}}\left(2 N_{t}-1\right) \beta_{0} \beta_{1} \beta_{2}(1-\alpha)\left(1-\alpha_{f_{1}}\right)\left(1-\alpha_{f_{2}}\right), \nonumber \\
				&+\frac{P^2}{4 N_{t}}\beta_{0}\left[2 N_{t} \beta_{1} \beta_{2}-\left(\alpha\left(\left|a_{00}\right|^{2}+\left|a_{11}\right|^{2}+\left|b_{00}\right|^{2}+\left|b_{11}\right|^{2}\right)+(1-\alpha)\left(\left|a_{10}\right|^{2}+\left|a_{01}\right|^{2}+\left|b_{10}\right|^{2}+\left|b_{01}\right|^{2}\right)\right)\right],  \\
				&d_{3}=P \beta_{0}+\frac{P^2}{4 N_{t}} \beta_{0}^{2}\left[N_{t}-2 \alpha+2 \alpha^{2}\right]-2 P \beta_{0}(1-\alpha)-\frac{P^2}{2 N_{t}}\beta_{0}^{2}\left(2 N_{t}-1\right)(1-\alpha)^{2}.
			\end{align}
		\end{subequations}
		
	\end{figure*}
	
	In Fig. \ref{fig5}, we plot the capacity versus RIS size for $P$ = 10 $\rm dB$ and $P$ = 12 $\rm dB$ to verify the derivation of (\ref{leq}). From this figure, it can be observed that the theoretical values derived from (\ref{leq}) closely match the simulated results, thereby confirming the accuracy of our derivation. Additionally, it is evident that the DP capacity increases more rapidly as the RIS size increases. This result is consistent with \textit{Remark 1}.
	
 \textcolor{black}{In Fig. \ref{fig4}, we plot the required RIS size versus the transmission power for different XPD values. The required RIS size decreases as $P$ increases since the full-rank advantage of the DP channel is exploited for larger $P$ levels. Furthermore, as $\alpha$ increases, indicating more severe polarization leakage, more elements are required because the interference between the two polarizations becomes more significant.
	Moreover, fewer elements are needed with fewer transmit antennas. This is because the third term in (11) becomes dominant when the number of transmit antennas is small and $\prod_{i=1}^2 \lambda_{i}$ in the DP system is larger than in the SP system, leading to a better DP performance.  Additionally, as $P \rightarrow \infty$, we see that the required number of elements reaches a lower limit.}

	\section{Conclusions}
	In this letter, we have derived the capacities of DP and SP MIMO systems. Based on these expressions, we have further derived the RIS size required for the DP MIMO system to outperform an equivalent SP system. From the analysis and simulation, we have the following conclusions: 1) The DP and SP capacities are proportional to the fourth and second power of the number of RIS elements, respectively. 2) The required size is a decreasing function of $P$ but there exists a lower bound for the required RIS size even with infinite power; 3) A larger RIS is required when the polarization leakage is more severe.
	
	\appendix 
	\subsection{Derivation of the channel capacity}\label{Apb}
	A closed-form expression for the capacity can be obtained by calculating $\sum_{i=1} \lambda_{i}$ and $\prod_{i=1} \lambda_{i}$ in (\ref {W}). Due to the properties
	\begin{equation}\label{lam}
		\begin{aligned}
			&	\sum_{i=1} \lambda_{i}=\rm {trace}\left(\mathbf{H} \cdot \mathbf{H}^{\mathrm{H}}\right),	\prod_{i=1} \lambda_{i}=\operatorname{det}\left(\mathbf{H} \cdot \mathbf{H}^{\mathrm{H}}\right),
		\end{aligned}
	\end{equation}
	we will derive $\operatorname{trace}\left(\mathbf{H} \cdot \mathbf{H}^{\mathrm{H}}\right)$ and $\operatorname{det}\left(\mathbf{H} \cdot \mathbf{H}^{\mathrm{H}}\right)$ for the DP and SP channels.
	
	\textit{ 1) DP channel capacity:} The reflected channel is denoted as
	\begin{equation}
		\begin{array}{l}
			\overline{\mathbf{H}}=\mathbf{H}_{2} \boldsymbol{\Phi} \mathbf{H}_{1} 
			=\left[\begin{array}{ll}
				\overline{\mathbf{h}}_{00} & \overline{\mathbf{h}}_{01} \\
				\overline{\mathbf{h}}_{10} & \overline{\mathbf{h}}_{11}
			\end{array}\right]^{2 \times 2 N_{t}},
		\end{array}
	\end{equation}
	where  ${\overline {\bf h}}_{pq} \in {\mathbb C}^{1 \times N_t}$ is the channel from polarization $q$ to polarization $p$, which can be rewritten as $
	\overline{\mathbf{h}}_{{pq}}=\left(a_{p q} x_{0}+b_{p q} x_{1}\right) \mathbf{a}_{N_{t}}^{\mathrm{T}}\left(\theta_{\mathrm{AoD}, 1}\right)$, where
	\begin{equation}
		\begin{array}{l}
			x_{0}={\bf a}_{L}^{\mathrm{T}}\left(\theta_{\mathrm{AoD}, 2}\right) {\bf \Phi}_{0} {\bf a}_{L}\left(\theta_{\mathrm{AoA}, 1}\right), \\
			x_{1}={\bf a}_{L}^{\mathrm{T}}\left(\theta_{\mathrm{AoD}, 2}\right) {\bf \Phi}_{1} a_{L}\left(\theta_{\mathrm{AoA}, 1}\right),
		\end{array}
	\end{equation}
	and $a_{pq}, b_{pq}$ are shown in (\ref {ab}). 
	\begin{figure*}[ht] 
		\centering 
		\vspace*{1pt} 
		\begin{equation}\label{ab}
			\begin{aligned}
				\begin{array}{l}
					a_{p q} \in \sqrt{\beta_{1} \beta_{2}}\left[\begin{array}{cc}
						\sqrt{1-\alpha_{f_{1}}} \sqrt{1-\alpha_{f_{2}}} e^{j\left(\psi_{1,00}+\psi_{2,00}\right)} & \sqrt{1-\alpha_{f_{1}}} \sqrt{\alpha_{f_{2}}} e^{j\left(\psi_{1,00}+\psi_{2,01}\right)} \\
						\sqrt{\alpha_{f_{1}}} \sqrt{1-\alpha_{f_{2}}} e^{j\left(\psi_{1,10}+\psi_{2,00}\right)} & \sqrt{\alpha_{f_{1}}} \sqrt{\alpha_{f_{2}}} e^{j\left(\psi_{1,10}+\psi_{2,01}\right)}
					\end{array}\right], \\
					b_{p q} \in \sqrt{\beta_{1} \beta_{2}}\left[\begin{array}{ll}
						\sqrt{\alpha_{f_{1}}} \sqrt{\alpha_{f_{2}}} e^{j\left(\psi_{1,01}+\psi_{2,10}\right)} & \sqrt{\alpha_{f_{1}}} \sqrt{1-\alpha_{f_{2}}} e^{j\left(\psi_{1,01}+\psi_{2,11}\right)} \\
						\sqrt{1-\alpha_{f_{1}}} \sqrt{\alpha_{f_{2}}} e^{j\left(\psi_{1,11}+\psi_{2,10}\right)} & \sqrt{1-\alpha_{f_{1}}} \sqrt{1-\alpha_{f_{2}}} e^{j\left(\psi_{1,11}+\psi_{2,11}\right)}
					\end{array}\right].
				\end{array}
			\end{aligned}
		\end{equation}
		\hrulefill 
	\end{figure*}
	Then, we calculate $\overline{\mathbf{h}}_{p q} \overline{\mathbf{h}}_{p q}^{\mathrm{H}}$, $\mathbb{E}\left\{\tilde{\mathbf{h}}_{p q} \tilde{\mathbf{h}}_{p q}^{\mathrm{H}}\right\}$, and $\mathbb{E}\left\{\tilde{\mathbf{h}}_{p q}^{\mathrm{H}} \tilde{\mathbf{h}}_{p q}\right\}$ in the following.
	\begin{align}
		&\overline{\mathbf{h}}_{p q} \overline{\mathbf{h}}_{p q}^{\mathrm{H}} \nonumber \\&=N_{t}\left(\left|a_{p q} x_{0}\right|^{2}+\left|b_{p q} x_{1}\right|^{2}+2 \operatorname{Re}\left\{a_{p q} b_{p q}^{*} x_{0} x_{1}^{*}\right\}\right), \label {zsd1}\\
		&\mathbb{E}\left\{\tilde{\mathbf{h}}_{p q} \tilde{\mathbf{h}}_{p q}^{\mathrm{H}}\right\} \in N_{t} \beta_{0}\left[\begin{array}{ll}
			1-\alpha & \alpha \\
			\alpha & 1-\alpha
		\end{array}\right], \label {zsd2}\\
		&\mathbb{E}\left\{\tilde{\mathbf{h}}_{p q}^{\mathrm{H}} \tilde{\mathbf{h}}_{p q}\right\} \in N_{t} \beta_{0}\left[\begin{array}{ll}
			(1-\alpha){\bf I}_{N_t} & \alpha{\bf I}_{N_t} \\
			\alpha{\bf I}_{N_t} & (1-\alpha){\bf I}_{N_t}
		\end{array}\right]. \label {zsd3}
	\end{align}

	By employing (\ref {zsd1}), (\ref {zsd2}), and (\ref {zsd3}) in (\ref {lam}), we obtain
	\begin{align}
		&\mathbb{E}\left\{\operatorname{trace}\left(\mathbf{H} \cdot \mathbf{H}^{\mathrm{H}}\right)\right\}= \nonumber \\
		&N_{t}\beta_{1} \beta_{2}(\left|x_{0}\right|^{2}+\left|x_{1}\right|^{2})
		+\operatorname{Re}\left\{r_{1} x_{0} x_{1}^{*}\right\}+2 N_{t} \beta_{0},\\
		&\mathbb{E}\left\{\operatorname{det}\left(\mathbf{H} \cdot \mathbf{H}^{\mathrm{H}}\right)\right\} 
		=\nonumber \\
		&N_{t}^{2}\left|a_{00} b_{11}+b_{00} a_{11}-a_{01} b_{10}-a_{10} b_{01}\right|^{2}\left|x_{0}\right|^{2}\left|x_{1}\right|^{2} \nonumber \\
		&	+N_{t}^2 \beta_{0}\beta_{1}\beta_{2}(\left|x_{0}\right|^{2} +\left|x_{1}\right|^{2})	+N_{t} \beta_{0}^{2}\left[N_{t}-2 \alpha+2 \alpha^{2}\right] \nonumber \\
		& -N_{t} \beta_{0}{\rm t}_0\left|x_{0}\right|^{2} -N_{t} \beta_{0}{\rm t}_1\left|x_{1}\right|^{2} 	+\operatorname{Re}\left\{r_{2} x_{0} x_{1}^{*}\right\},
	\end{align}
	where 
	\begin{align}
		&r_{1}=2 N_{t}\left(a_{00} b_{00}^{*}+a_{11} b_{11}^{*}+a_{10} b_{10}^{*}+a_{01} b_{01}^{*}\right), \\
		&r_{2}=2 N_{t} \beta_{0} [N_{t}(a_{00} b_{00}^{*}+a_{11} b_{11}^{*}+a_{10} b_{10}^{*}+a_{01} b_{01}^{*})   \nonumber  \\
		&-\left(\alpha(a_{00} b_{00}^{*}+a_{11} b_{11}^{*})+(1-\alpha)(a_{10} b_{10}^{*}+a_{01} b_{01}^{*})\right)], \\
		&{\rm t}_0=\alpha(|a_{00}|^{2}+|a_{11}|^{2})+(1-\alpha)(|a_{10}|^{2}+|a_{01}|^{2}), \\
		&{\rm t}_1=\alpha(|b_{00}|^{2}+|b_{11}|^{2})+(1-\alpha)(|b_{10}|^{2}+|b_{01}|^{2}).
	\end{align}
	
	\textit{ 2) SP channel capacity:} The reflected channel can be expressed as
	$\overline{\mathbf{h}}_{p q}=\left(c_{p q} x\right) \mathbf{a}_{N_{t}}\left(\theta_{\mathrm{AoD}, 1}\right)$,
	where $x={\bf a}_{2 L}^{\mathrm{T}}\left(\theta_{\mathrm{AoD}, 2}\right) {\bf \Phi} {\bf a}_{2 L}\left(\theta_{\mathrm{AoA}, 1}\right)$. The coefficients $c_{pq}$ are omitted here since they can be easily calculated.
	Then, by calculating $\overline{\mathbf{h}}_{p q} \overline{\mathbf{h}}_{p q}^{\mathrm{H}}$, $\mathbb{E}\left\{\tilde{\mathbf{h}}_{p q} \tilde{\mathbf{h}}_{p q}^{\mathrm{H}}\right\}$, and $\mathbb{E}\left\{\tilde{\mathbf{h}}_{p q}^{\mathrm{H}} \tilde{\mathbf{h}}_{p q}\right\}$ and employing them in (\ref {lam}), we obtain
	\begin{align}	
		&\mathbb{E}\left\{\operatorname{trace}\left(\mathbf{H} \cdot \mathbf{H}^{\mathrm{H}}\right)\right\}= \nonumber \\
		&4 N_{t} \beta_{0}(1-\alpha)+4 N_{t} \beta_{1}\beta_{2}\left(1-\alpha_{f_{1}}\right)\left(1-\alpha_{f_{2}}\right)|x|^{2},\\
		&\mathbb{E}\left\{\operatorname{det}\left(\mathbf{H} \cdot \mathbf{H}^{\mathrm{H}}\right)\right\}= 2\left(2 N_{t}^{2}-N_{t}\right) \beta_{0}^{2}(1-\alpha)^{2}+\nonumber \\
		&4\left(2 N_{t}^{2}-N_{t}\right) \beta_{0} \beta_{1} \beta_{2}(1-\alpha)\left(1-\alpha_{f_{1}}\right)\left(1-\alpha_{f_{2}}\right)|x|^{2}.
	\end{align}
	
	\textit{ 3) Phase shift design:} The capacity can be maximized by optimizing the phase shift matrix of the RIS \cite{9491943}. 
	For (\ref {W}) applied to the DP MIMO case,  $|x_0|^2, |x_1|^2$ are maximized when the phase shifts in its polarization are matched,  and  ${\rm Re}\{r(P)x_0 x_1^*\}$ is maximized if $r(P)x_0 x_1^*$ is real, i.e., $\angle r+x_0-x_1=0$ \cite {9497725}. Thus, \textcolor{black} {the upper bound for the capacity is obtained when $\theta_{0}-\theta_{1}=-\angle r(P),$}
	\begin{align}
		&\phi_{0, n}+2 n \pi \frac{d}{\lambda}\left(\sin \theta_{\mathrm{AoA}, 1}+\sin \theta_{\mathrm{AoD}, 2}\right)=\theta_{0}, \\
		&\phi_{1, n}+2 n \pi \frac{d}{\lambda}\left(\sin \theta_{\mathrm{AoA}, 1}+\sin \theta_{\mathrm{ADD}, 2}\right)=\theta_{1}, 
	\end{align}
	where $r(P)=\frac{P^{2}}{4 N_{t}^{2}} r_{2}+\frac{P}{2 N_{t}} r_{1}$. For the SP MIMO case, the approximate phase shift design is $\phi_{n}+2n\pi \frac{d}{\lambda}\left(\sin \theta_{\mathrm{AoA}, 1}+\sin \theta_{\mathrm{AoD}, 2}\right)=\theta.$ \textcolor{black}{The upper bound for DP and SP capacity are then obtained using the expression in (\ref {DP_op}) and (\ref {SP_op}), respectively.}
	
	\subsection{A proof of inequality (\ref {161}) }\label{Apa}
	The coefficient is written as $d_{3}=P \beta_{0}(2\alpha-1) +\frac{P^2}{4 N_{t}} \beta_{0}^{2} g(\alpha)$,	where $g(\alpha)=\left[ N_{t}-2 \alpha+2 \alpha^{2}-2\left(2 N_{t}-1\right)(1-\alpha)^{2} \right]$. Then, since $N_t>=1$ and $\alpha \in (0,1]$, we have	$g^{\prime}(\alpha)=(1-N_t)\alpha+N_t-\frac{3}{4} \ge \frac{1}{4}.$ Thus, $g(\alpha)$ is an increasing function of $\alpha \in (0,1]$ and its zero point is $1/2$. Then we obtain $d_3<0$ if $\alpha <1/2$.
	\bibliographystyle{IEEEtran}
	\bibliography{document.bib}
\end{document}